\documentclass[aps,prd,a4paper,floats,nofootinbib]{revtex4} 
\usepackage{graphicx}
\usepackage{subfigure}
\usepackage{float}

\begin{document}

\title{Doppler magnification in flux-limited galaxy number counts with finite redshift bin width}
\author{
Song Chen$^{a,b}$\\~
\\
\emph{$^a$Department of Physics \& Astronomy, University of the Western Cape,
Cape Town 7535, South Africa \\
$^b$Amalienstra{\ss}e 38, M{\"u}nchen 80799, Germany\\
}}

\date{\today}

\begin{abstract}
In this paper, I investigated the Doppler magnification effect
in the flux-limited galaxy number counts with finite redshift bin width.
In contradiction with our intuition, the number counts correction formula for different redshift bin are different.
For the redshift window as delta function and constant function, the correction formulas have been derived analytically in this paper. 
These two windows correspond to two extreme cases(i.e. extremely narrow bin and extremely broad bin).
An simulation have been implemented to test these formulas.
The simulation results indicate the perturbation 
changes smoothly from one extreme case to another while extending the redshift bin width from half bin width $0.05$ to $0.4$.
As a result, Doppler magnification caused number counts perturbation of finite redshift bin width can NOT be compute via redshift integration of the perturbation formula derived with delta function redshift window.
It is only a good approximation when the bin width is small.
These results are important for galaxy redshift survey kinematic dipole estimation, and meaningful for the relativistic galaxy number counts angular power spectrum estimation, such as CLASSgal\cite{2013JCAP...11..044D} and CAMB source\cite{Challinor:2011bk}.
\end{abstract}

\maketitle
\section{\label{sec:1} Introduction}

In galaxy surveys, observers count the number of galaxies per fixed solid angle in direction $\mathbf{n}$ within certain redshift range $z_{min}\to z_{max}$. Then, they average the number counts over directions to obtain the 'mean' number counts of galaxies per fixed solid angle per fixed redshift range.
Based on above information, cosmologists attempt to learn about the formation and evolution of large scale structures.
However, comparing the measurements with the theoretical predictions for the cosmological models, requires 'coordinates transformation'.
This transformation helps changing the theoretical spatial coordinates into the observed coordinates.
Many perturbations other than the intrinsic cosmological matter density perturbation are involved as byproducts of the this transformation. 
Doppler magnification is one of them.

In the real observation, some galaxies look brighter than they should be due to their peculiar motions w.r.t us. This effect is named Doppler magnification \cite{2008PhRvD..78l3530B,2014MNRAS.443.1900B}. 
Considering the fact that our galaxy redshift surveys are flux or magnitude limited, Doppler magnification do perturb the observed galaxy counts through the observed flux or magnitude cutoff.

The fluctuation of the observed galaxy number counts per solid angle consist of two big contributions.
One is the solid angle area $\Omega$ fluctuation due to the effects such as gravitational lensing and relativistic aberration. 
The other contribution is from the galaxy surface density $\rho$ fluctuation which caused by the effects such as underlying matter density fluctuation $\delta_g$, volume distortion along the line of sight as while as the Doppler magnification.

The previous studies on the galaxy number counts relativistic corrections are focus on the infinitely narrow redshift bin width cases\cite{2009PhRvD..79b3517Y,2011PhRvD..84f3505B,Challinor:2011bk,2012PhRvD..85b3504,2014JCAP...09..037B}. To be more specific, the Dirac delta function $\delta(z-z_*)$ redshift window has been assumed in their derivations.
Given a finite redshift bin width, the expected perturbations of this bin are commonly estimated via redshift integration of the perturbation formula derived in delta function redshift window case.  
However, this method haven't been questioned before.

This paper is focusing on investigating the connection between
the Doppler magnification effect caused number counts perturbation and number counts redshift bin width. 
I first review a few aspects about the peculiar motion perturbations in redshift and luminosity distance. In section \ref{sec:3}, I show the standard method to derive the surface density perturbation caused by flux fluctuations. Following this section, two extreme redshift window functions have been investigated in section \ref{sec:4} and \ref{sec:5}.
In order to test my arguments, A series of number counts simulations have been implemented in section \ref{sec:6}.

\section{\label{sec:2} Doppler magnification}
Before I discuss the number counts perturbation caused by Doppler magnification effect, it is necessary to review a few important aspects about redshift, luminosity distance and Doppler magnification.

First, the observed redshift may not straightly reflect the source underlying distance. More often, the observed redshifts are affected by the Doppler effect, Sachs-Wolfe effect as well as integrated Sachs-Wolfe effect.

The redshift perturbations are most likely dominated by the peculiar velocity at the observed galaxy $\mathbf{v}$ and observer $\mathbf{v}_o$. 
Approximately, the observed redshift can be written as
\begin{eqnarray}
 1+z=\frac{1}{a(\eta)}(1 + \mathbf{n}\cdot\mathbf{v}-\mathbf{n}\cdot\mathbf{v}_o)=(1+\bar{z})(1+\delta z)\;,
\end{eqnarray}
where $a$ is the scale factor, and the scale factor today is $a_0=1$. 
$\bar{z}={1}/{a(\eta)}-1$ is the background redshift, and $\mathbf{n}$ is the unit vector from observer to emitter.

Similar to redshift, luminosity distance also affect by the peculiar motion at the observer and source. This is the Doppler magnification effect.
The formula of luminosity distance up to linear order\cite{1987MNRAS.228..653S} has been derived many times.
Here, I skip the full derivation and quote the formula with relevant linear perturbations directly,
\begin{eqnarray}\label{eq:lumdist}
 d_L&=& (1+\bar{z})r(\bar{z})(1+2\mathbf{n}\cdot\mathbf{v}-\mathbf{n}\cdot\mathbf{v}_o)\equiv\bar{d}_L(\bar{z})(1+\delta d_L^{\bar{z}})\nonumber\\
 d_L&=&(1+z) r(z)(1+\mathbf{n}\cdot\mathbf{v}-\frac{1}{\mathcal{H}(z) r(z) }(\mathbf{n}\cdot\mathbf{v}-\mathbf{n}\cdot\mathbf{v}_o))\equiv\bar{d}_L(z)(1+\delta d_L^{z})
\end{eqnarray}
where $\mathcal{H}$ is conformal Hubble parameter, and $r(z)$ is the comoving distance evaluated at observed redshift $z$.
It is clear to see that the luminosity distance fluctuations are based on the choice of redshift(background redshift $\bar{z}$ or observed redshift $z$).

Considering the fact that flux $F$ can be written as a ratio between luminosity and luminosity distance square,
\begin{eqnarray}
F=\frac{L}{4\pi d_L^2}\;,
\end{eqnarray}
the flux fluctuations are based on the choice of redshift as well.
Doppler magnification effect modifies the observed flux of each individual galaxy. Suppose an galaxy survey can observe infinitely faint galaxy, then the Doppler magnification do not contribute to the observed galaxy number counts fluctuation.
To affect the observed galaxy number counts, A number counting associated with flux selection is necessary. 

\section{\label{sec:3} The Doppler magnification effect in the flux limited observed galaxy number counts}

All galaxy surveys are limited by the flux sensitivity of the telescope.
Considering the fact that only bright sources are observed, 
the galaxy surface density $\rho$ of the given redshift bin can be considered as an integration of luminosity function $f(z,L)$ and volume factor $\frac{r^2}{\mathcal{H}}$ over redshift and luminosity,
\begin{equation}
 \rho(\mathbf{n},F_*)=\int_0^\infty {\rm d}z W(z) \frac{r^2}{\mathcal{H}}\int^\infty_{4\pi F_* d_L^2 } {\rm d}L \;f(z, L)
\end{equation}
where $W(z)$ is redshift window of the given redshift bin, $F_*$ is the flux limit of the survey, and $d_L$ is the luminosity distance\footnote{Here, I ignore all perturbations}.

Although the observed flux limit in all direction are same, the
luminosity cutoff $4\pi F_* d_L(\mathbf{n},z)^2$ is not isotropic
due to the existence of luminosity distance fluctuations.
This is exactly how Doppler magnification contribute to galaxy number counts fluctuation.
It is important to point out, there is an nature isotropic luminosity cutoff. Assuming the luminosity function is same everywhere, 'no luminosity cutoff' is a 'isotropic luminosity cutoff'\footnote{Here, I do not consider the matter density fluctuation and bias.}.
As a result, the galaxy surface density perturbation of isotropic flux limited survey for a given redshift bin equals to the galaxy surface density perturbation without flux limit for the same redshift bin.

\begin{figure}
  \centering
  \subfigure[]{\includegraphics[width=0.4\textwidth]{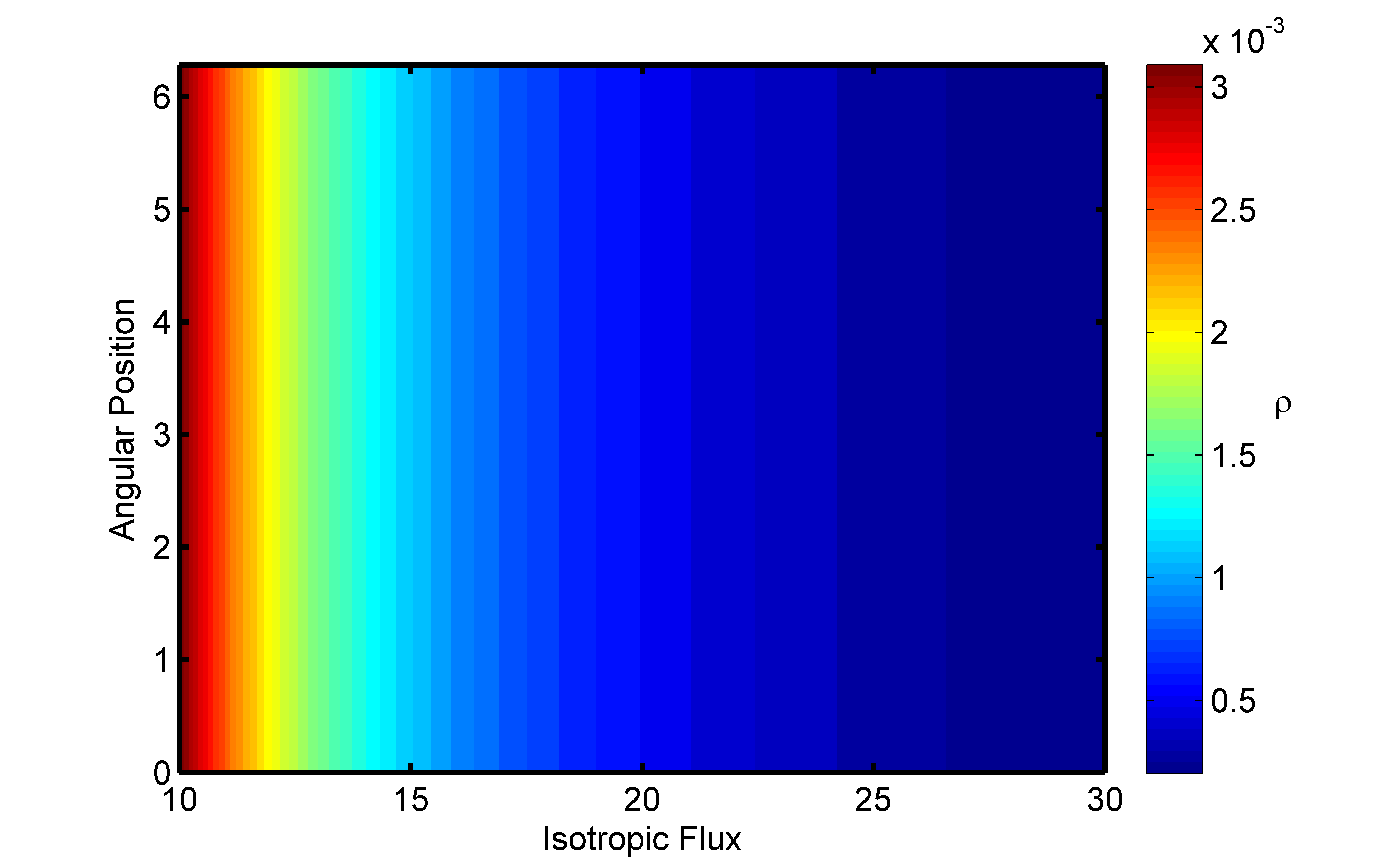}\label{fig:iso_F}}
  \subfigure[]{\includegraphics[width=0.4\textwidth]{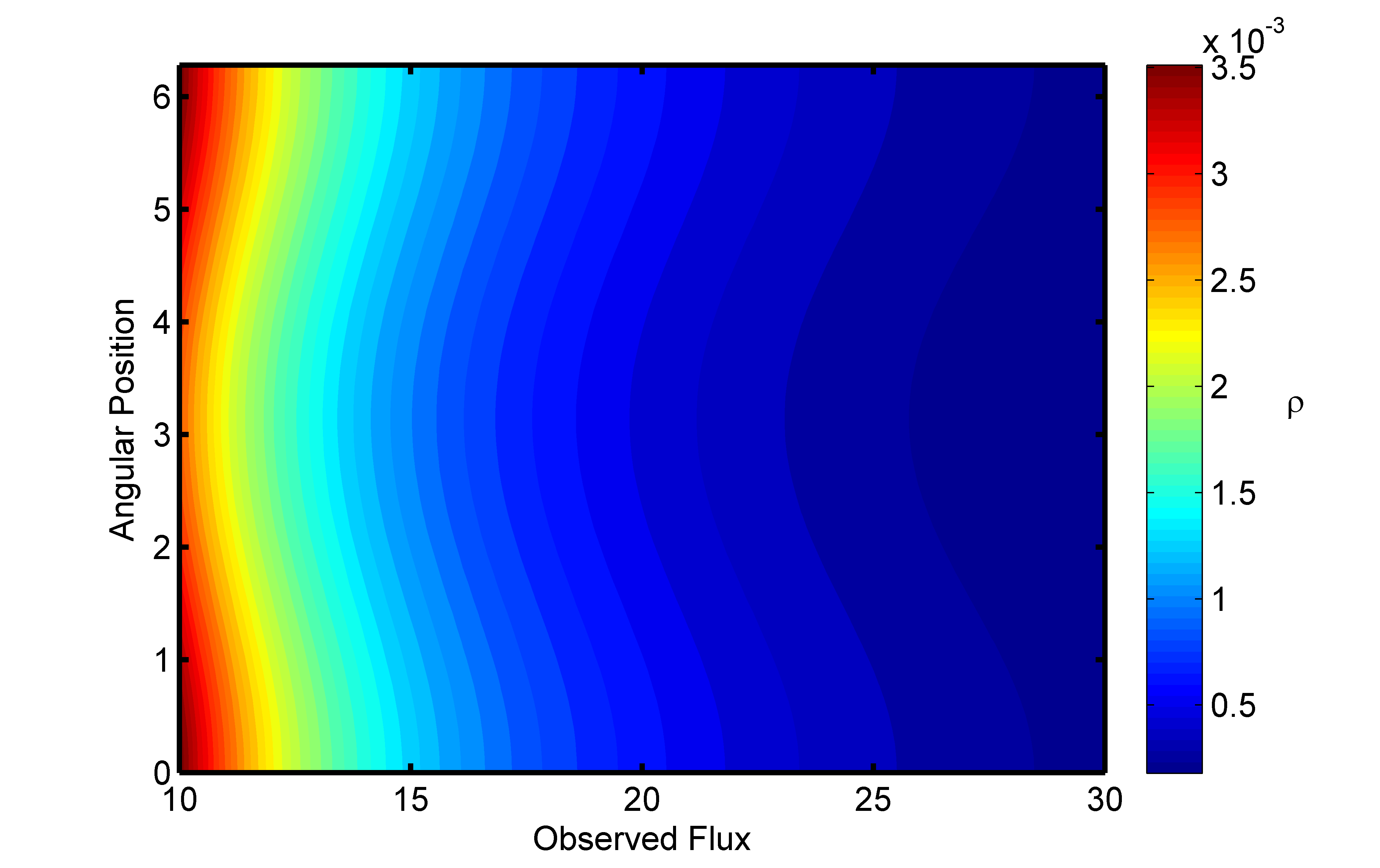}\label{fig:obs_F}}
  \caption{Schematic diagram for density perturbation caused by different coordinates: isotropic flux vs observed flux. The vertical axis is one dimensional angular position from $0$ to $2\pi$. The two figures are corresponding to one realization, in which the galaxy surface density $\rho$ is set to be isotropic(keeping constant along the angular position axis) in the isotropic flux $\tilde{F}$.
The surface density fluctuation for a given flux in the right figure is caused by the mapping from $F\to\tilde{F}$ involved angular position.}
  \label{fig:slava}
\end{figure}

In principle, it is possible to find an isotropic flux $\tilde{F}$, which guarantee isotropic luminosity cutoff.
This isotropic flux $\tilde{F}$ is NOT necessarily equal to the flux observed in the background universe $\bar{F}$(This is why we using tilde instead of bar for the isotropic flux).
The direct measurement, observed flux $F$, relate to the isotropic flux $\tilde{F}$ via $F=\tilde{F}+\delta F$. 
Here, we consider $F$ as new coordinate, and $\rho$ on the hyper-surface $F=\text{const.}$ depends on the angular position $\mathbf{n}$ (see Figure \ref{fig:slava}).
Ignoring the intrinsic cosmological galaxy density fluctuation, we have
\begin{equation}
 \rho(\tilde{F})=\rho(F-\delta F)\approx \rho(F)-\frac{\partial \rho}{\partial F}\delta F\equiv\bar{\rho}(F)+\delta \rho(F,\mathbf{n})
\end{equation}
where $\bar{\rho}(F)$ is the background galaxy surface density in the observed flux $F$ coordinate, while $\delta \rho(F,\mathbf{n})$ describes linear surface density perturbation.
Similar to the gauge dependent metric perturbations, this perturbation is gauge(or coordinate) dependent. It depends on how we choose flux.
Contradict to the gauge dependent metric perturbations, the gauge choice of this perturbation is fixed by the observer, i.e given one observer, there is only one observed flux $F$. 
The flux-limited galaxy surface density fluctuation is
\begin{equation}
 \delta \rho(F,\mathbf{n})=-\frac{\partial \bar{\rho}}{\partial F}\delta F
\end{equation}
In the previous studies\cite{2013JCAP...11..044D,Challinor:2011bk}, many papers use luminosity instead of flux as coordinate. 
Here, we use flux because it is the direct observable. 
The final result is not affect by this preference.  
Because luminosity distance is the core of the whole method.

By combining background luminosity distance with various perturbations, we create different 'flux' coordinates respectively.
The key issue to estimate the galaxy surface density fluctuation is to find the correct isotropic flux $\tilde{F}$ as the expansion flux. 
However, this task is not as trivial as it looks like. 
Since the luminosity distance is a function of redshift,
the redshift window function affects the luminosity distance distribution.
This fact further affects the choice of isotropic flux for the given bin.


In the following sections, two specific redshift window functions will be discussed. They are corresponding to two extreme cases, i.e. infinitely narrow bin and infinitely broad bin. 

\section{ \label{sec:4} Observed Redshift Window: Delta Function }
One of the simplest redshift window function is delta function, $W(z)=\delta(z-z_*)$. 
Since the window is acting on the observed redshift, all the galaxies in this bin have the same observed redshift $z_*$. 
Consequently, the background luminosity distances $\bar{d}_L(z)$ evaluated at observed redshift $z_*$ are same for all galaxies inside the bin regardless their angular positions.
Therefore, the galaxy surface density with redshift window $\delta(z-z_*)$ and flux limit $L/(4\pi \bar{d}_L(z)^2 )>F_*$,
\begin{eqnarray}
 \tilde{\rho}(z_*,F_*)&=&\int_0^\infty {\rm d}z \delta(z-z_*)  \frac{r^2}{\mathcal{H}}\int^\infty_{4\pi F_* \bar{d}_L(z)^2 } {\rm d}L \;f(z, L)\nonumber\\
 &=&\frac{r^2(z_*)}{\mathcal{H}(z_*)}\int^\infty_{4\pi F_* \bar{d}_L(z_*)^2 } {\rm d}L \;f(z_*, L)\;,
\end{eqnarray}
is direction independent.  
Thus, the isotropic flux in this redshift window is 
\begin{equation}\label{eq:flux1}
 \tilde{F}=\frac{L}{4\pi \bar{d}_L(z)^2}\;.
\end{equation}
Combining Eq.~\ref{eq:lumdist} and Eq.~\ref{eq:flux1},
the corresponding flux fluctuations due to Doppler magnification is
\begin{eqnarray}
 \frac{\delta F}{\tilde{F}}&=&(\frac{\bar{d}_L(z)}{ d_L(z)})^2-1=-2[\mathbf{n}\cdot\mathbf{v}-\frac{1}{\mathcal{H}(z) r(z)}(\mathbf{n}\cdot\mathbf{v}-\mathbf{n}\cdot\mathbf{v}_o)]\;.
\end{eqnarray}
Thus, the Doppler magnification caused surface density fractional perturbation $\Delta_N$(i.e. $\Delta_N\equiv{\delta\rho}/\bar{\rho}$) becomes
\begin{eqnarray}\label{eq:delta_narrow}
 \Delta_N(\mathbf{n},z, F)&=&\Delta_N(\mathbf{n},z)|_{\tilde{F}}-\frac{\partial \ln \rho}{\partial\ln F}\frac{\delta F}{F}\nonumber\\
 &=&\Delta_N(\mathbf{n},z)|_{\tilde{F}}+\frac{\partial \ln \rho}{\partial\ln F}[2-\frac{2}{\mathcal{H}(z) r(z)}]\mathbf{n}\cdot\mathbf{v}+\frac{\partial \ln \rho}{\partial\ln F}\frac{2}{\mathcal{H}(z) r(z)}\mathbf{n}\cdot\mathbf{v}_o
\end{eqnarray}

In the optical and infrared galaxy surveys, magnitude is often used instead of flux.
It is convenient to introduce magnification bias $s$,
\begin{eqnarray}
\frac{\partial \ln \rho}{\partial\ln F}=-\frac{5}{2} s
\end{eqnarray}
It is worth to clarify that the magnification bias $s$ depends on the redshift window as well. 
Therefore, Eq.~\ref{eq:delta_narrow} can be rewrite as
\begin{eqnarray}\label{eq:win_delta}
 \Delta_N(\mathbf{n},z, F)&=&\Delta_N(\mathbf{n},z)|_{\tilde{F}}-[5s-\frac{5s}{\mathcal{H}(z) r(z)}]\mathbf{n}\cdot\mathbf{v}-\frac{5s}{\mathcal{H}(z) r(z)}\mathbf{n}\cdot\mathbf{v}_o\;.
\end{eqnarray}
The second term in the formula agrees with previous studies\cite{Challinor:2011bk,2013JCAP...11..044D}.
The last term in the formula agrees with our previous study about the kinematic dipole amplitude of HI galaxy \cite{2018JCAP...01..013M}.
Note, my definition for $\mathbf{n}$ is different from
the definition in papers\cite{2006PhRvD..73b3523B,2013JCAP...11..044D} for a minus sign.

\section{ \label{sec:5} Observed Redshift Window: Constant Function }
The other extreme window function is constant function, i.e. $W(z)=1$. 
This case corresponding to projected or continuum galaxy number counts analysis. 
Different from the above case, neither observed redshift $z$ nor background redshift $\bar{z}$ is fixed by the redshift window function. 
To find the isotropic flux for this window function, one has to relying on the fundamental assumption in cosmology that
the universe is statistic isotropic and homogeneous.
Thus, the galaxy surface density should be isotropic in the background universe(without peculiar motion and metric perturbations).
Based on this logic, the isotropic flux $\tilde{F}$ equals to the flux observed in the background universe,
\begin{equation}\label{eq:flux2}
 \tilde{F}=\frac{L}{4\pi \bar{d}_L(\bar{z})^2}\;,
\end{equation}
and the surface density
\begin{eqnarray}
 \tilde{\rho}(F_*)&=&\int_0^\infty {\rm d}z \frac{r^2}{\mathcal{H}}\int^\infty_{4\pi F_* \bar{d}_L(\bar{z})^2 } {\rm d}L \;f(z, L)\;,
 \end{eqnarray}
 is isotropic.
Combining Eq.~\ref{eq:lumdist} with Eq.~\ref{eq:flux2}, the Doppler magnification caused surface density fractional perturbation for the projected galaxy number counts is
\begin{eqnarray}\label{eq:delta_broad}
 \Delta_N(\mathbf{n},F)=4\frac{\partial \ln \rho}{\partial\ln F}\mathbf{n}\cdot\mathbf{v}-2\frac{\partial \ln \rho}{\partial\ln F}\mathbf{n}\cdot\mathbf{v}_o =-10s\;\mathbf{n}\cdot\mathbf{v}+5s\;\mathbf{n}\cdot\mathbf{v}_o\;.
\end{eqnarray}
This formula has been derived before.

In the CLASSgal and CAMB source, their number counts perturbations are
evaluated for delta window function case.
They integrate the those perturbations over redshift as an approximation for the perturbations in galaxy number counts with finite redshift bin width.
It is important to point out that, this integration method
does not lead Eq.~\ref{eq:delta_narrow} to Eq.~\ref{eq:delta_broad}.
In another words, this method can not give correct Doppler magnification perturbations galaxy number counts with finite redshift bin width.
Because, the perturbation derivation are based on the expansion.
Extending the redshift window changes the 'benchmark' of the expansion(i.e. changes the isotropic flux from Eq.~\ref{eq:flux1} to Eq.~\ref{eq:flux2}).
However, the redshift window integration method assumes the same the isotropic flux, i.e. Eq.~\ref{eq:flux1}. 
Doppler magnification is not the only perturbation affect by the redshift bin width.
All the redshift fluctuations contained in the luminosity distance fluctuations have the same issue.

\section{\label{sec:6} Numerical Simulations}
In this section, I investigate the relation between the redshift bin width and isotropic flux with numerical simulations.
The propose of this simulation is to show different redshift bin has different the isotropic flux. 

In the simulation, I assume the galaxy number density is exactly same in all direction. For simplicity, I set the galaxy background redshift $\bar{z}$ following Gaussian distribution with mean $1.0$ and sigma $0.1$. 
In order to see the purely surface density perturbation due to the anisotropic luminosity cutoff, I let the galaxy luminosity following uniform distribution.
In all angular position, my sample is consist of $1,000,000$ background redshifts and luminosities drawn from distributions mentioned above. 

To test anisotropy caused by Doppler magnification,
I randomly draw $100$ redshift fluctuations $\delta z$ for $100$ different angular positions. 
At each angular position, $\delta z$ is same for all galaxies.
While $\delta z$ is generated via
\begin{equation}
\delta z=0.0123(2r-1)
\end{equation}
where $r\in [0,1)$ is a uniform distributed random number, and $0.0123$ is $10$ times of our peculiar motion respect to the CMB rest frame\cite{COBE:1996,Aghanim:2013suk}\footnote{The reason to generate perturbations in this way is to mimic the peculiar velocity of observer, where $2r-1$ gives $\cos(\theta)\in[-1,1)$.
Actually, $\theta$ is the one dimensional angular position I used in the simulation. }.
The simulated observed redshift is
\begin{eqnarray}
z=(1+\bar{z})(1+\delta z)-1\;.
\end{eqnarray}

After that, we adopt the observed redshift selection, i.e. $1.0-w <z<1.0+w$, where $w$ is the bin half width. This selection keeps the bin mean observed redshifts same for different $w$.  
\begin {table}
\begin{center}
\begin{tabular}{| c | c | c | }\hline
Group & flux limit & observed redshift selection\\ \hline
A & $L/ (4\pi\bar{d}_L(z)^2) > F_{*}$ & $1.0-w <z<1.0+w$  \\  \hline
B & ${L}/(4\pi\bar{d}_L(\bar{z})^2) > F_{*}$ & $1.0-w <z<1.0+w$  \\  \hline
C &  none & $1.0-w <z<1.0+w$ \\ \hline
\end{tabular}
\end{center}
\caption {Simulation counting group setting}
\label{tab:sim_group}
\end{table}
As shown in Table~\ref{tab:sim_group}, I count the galaxy number from three(selection) groups.
In group A, I set flux limit 
$L/ (4\pi\bar{d}_L(z)^2) > F_{*}$ following Eq.~\ref{eq:flux1} \footnote{I choose proper $F_{*}$ to allow enough galaxies pass the flux selection, based on the luminosity and redshift distribution. I keep $F_*$ invariant for different $w$ in the simulation. }.
In group B, I set flux limit 
${L}/(4\pi\bar{d}_L(\bar{z})^2) > F_{*}$ following Eq.~\ref{eq:flux2}.
I do not set any flux limit for group C.
As discussed in section~\ref{sec:3}, the fluctuation of the real isotropic flux limit galaxy count should equal to the fluctuation of no flux limit galaxy count. 
Considering  the fact that redshift perturbation also affect the galaxy number counts, I also need Group C to help us isolate the pure Doppler magnification effect.

To tell which isotropic flux matches with which redshift bin width,
an anisotropy estimator is needed to quantify the number counts anisotropic fluctuation.
If the anisotropic fluctuation amplitude of group A or B agree with group C,
then the corresponding flux cutoff used in group A or B is an isotropic flux cutoff.

\begin{figure}
  \centering
  \includegraphics[width=0.8\textwidth]{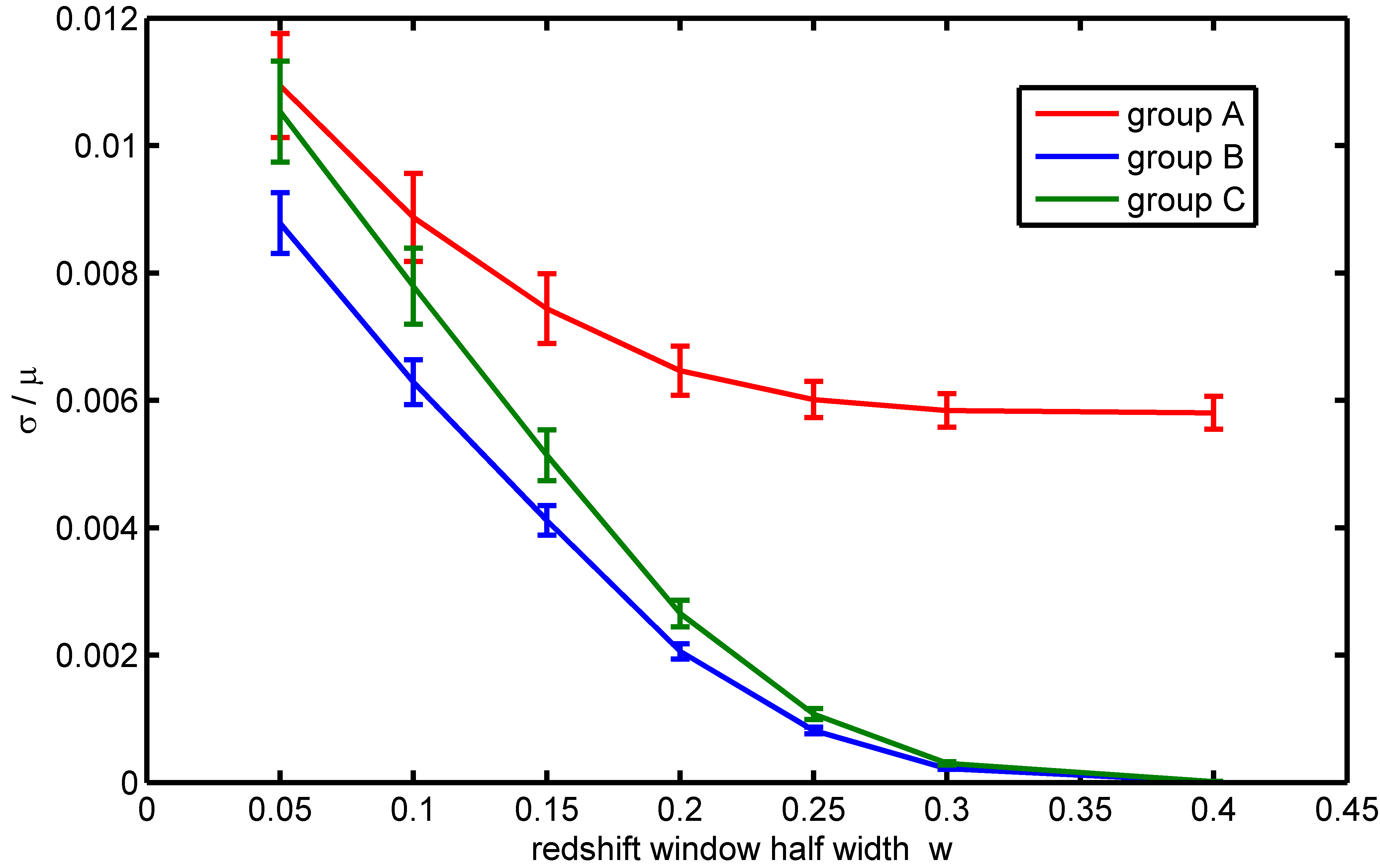}
  \caption{ Number counts standard deviation over mean vs. different redshift bin width: Group A, we set flux limit 
$L/ (4\pi\bar{d}_L(z)^2) > F_{*}$; Group B, we set flux limit 
${L}/(4\pi\bar{d}_L(\bar{z})^2) > F_{*}$; Group C, we set no flux limit. 
All the group adopted observed redshift selection: $1.0-w<z<1.0+w$. 
  }
  \label{fig:sim}
  \end{figure}

Hence, I calculate the mean and standard deviation of the galaxy number counts of the $100$ different angular positions for each group.
The ratio between standard deviation and mean(i.e. $\sigma/\mu=\sqrt{<\Delta_N^2>}$ where $\Delta_N$ is the surface density fractional perturbation) tells us how anisotropic the galaxy number counts is. 
For every $w$, the simulation has been repeated for $1,200$ times.
I fit the $1,200$ measurements of $\sigma/\mu$ with Gaussian distribution, and plot the fitted mean and sigma for each $w$ in Fig.~\ref{fig:sim}.

As shown in Fig.~\ref{fig:sim}, the group A and group C are agree with each other within $1$ sigma error at small $w$(i.e. at narrow redshift bin limit).
The deviation between A and C getting larger and larger as $w$ increasing. Finally, the group C and group B are overlapping for $w > 0.3$.
This result can be interpreted as following: 
When the observed redshift window is narrow, the isotropic flux for delta function case is a good approximation. The observed number counts fluctuation is purely due to the observed redshift perturbation and selection.
When the observed redshift window is broad enough, the isotropic flux for constant function case is a good approximation.  
We could see this isotropic flux curve asymptotic close to curve without flux limit. For group B and C, the number counts fluctuations turn to vanish as $w$ increasing. This is because the broad redshift selection includes almost all the galaxies and the redshift perturbation become negligible. When the redshift perturbation vanish, the group A measures purely Doppler magnification caused number counts fluctuation. This perturbation proportion to the deviation between the two isotropic flux(Eq.~\ref{eq:flux1} and Eq.~\ref{eq:flux2}).

When the observed redshift window is in-between, 
the isotropic flux of the two extreme cases do not fit the real isotropic flux(i.e. group C).
I haven't found the analytical expression for the intermediate isotropic flux.
It is clear that the curve of group C is in the middle of the curve of group A and B.
Therefore, the in-between isotropic flux $\tilde{F}^{in}$ must be between $L/ (4\pi\bar{d}_L(z)^2)$ and $L/ (4\pi\bar{d}_L(\bar{z})^2)$.

Moreover, since the redshift perturbation setting in the simulation is to mimic the observer's peculiar motion, the result of this simulation directly indicates that the measured galaxy number counts kinematic dipole amplitudes are different for different redshift bin width.

\section{\label{sec:7} Discussion}
In the previous study about the kinematic dipole\cite{2012MNRAS.427.1994G}, people found the measured 2MASS redshift survey\cite{2012ApJS..199...26H} kinematic dipole amplitude decreases from $0.275$ to $0.125$ as the redshift cutoff $z_{max}$ shifting from $0.02$ to $0.1$. 
According our previous paper\cite{2018JCAP...01..013M}, we know the kinematic dipole amplitude $D_{gal}$ for infinitively narrow redshift bin is
\begin{eqnarray}
D_{gal}=[3+\frac{\dot{H}(z)}{H(z)^2}+\frac{2-5s}{\mathcal{H}(z)r(z)}-b_e(z)]\mathbf{v}_o
\end{eqnarray}
where $b_e$ is the galaxy evolution bias, and $-5s\mathbf{v}_o/(\mathcal{H}(z)r(z))$
is the Doppler magnification effect same as Eq.~\ref{eq:win_delta}.
From our simulation, we can infer that the dipole contribution from Doppler magnification $D_{mag}$ will change from $-5s\mathbf{v}_o/(\mathcal{H}(z)r(z))$ to $5s \,\mathbf{v}_o$ by extending the $z_{max}$.
It is clear that the two perturbations have different sign.
The magnification bias $s$ is positive for 2MASS redshift survey\cite{2012ApJS..199...26H}.
Therefore, I expect to see the measured dipole amplitude increase as $z_{max}$ change from $0.02$ to $0.1$. 
However, we should also keep in mind that as long as we extend 
$z_{max}$, the mean redshift also increase. 
This shift of mean redshift causes the dominate term 
${2}/{\mathcal{H}(z)r(z)}$ decreasing. 
Thus, the measured dipole amplitude should be affect by these two effects.

The above investigation assumes the measured dipoles are completely due to the peculiar motion of the observer $\mathbf{v}_o$.
The current dipole direction measurements\cite{2012MNRAS.427.1994G} do not support with this assumption. Different measured dipole direction for different redshift bin suggest that there may be several coherent bulk flows in the redshift range of 2MASS, which amplified the contribution from source velocity $\mathbf{v}_s$. 
To verify the theory, A complete local galaxy bulk flow catalogue and next generation large sky coverage redshift surveys are needed. 


\section{\label{sec:8} Conclusion}
It is commonly believed that relativistic corrections for the galaxy number counts with finite redshift bin width 
can be compute via redshift integration of the those corrections derived in the infinitely narrow bin case.
In this paper, I illustrate an counter example, i.e. Doppler magnification effect.
The number counts correction formula for Doppler magnification effect
depends on the redshift window. 

For the narrow bin galaxy number counts, the correction to the angular power spectrum caused by the Doppler magnification from the source velocity $\mathbf{v}_s$ is not as significant as redshift space distortion\cite{2011PhRvD..84f3505B}. 
For the broad bin galaxy number counts, both of them are negligible.
Because the redshift selection effect is minimized in the broad bin.  

However, the investigation about the Doppler magnification from the observer velocity $\mathbf{v}_o$ in flux-limited galaxy number counts with finite redshift bin width helps improving the kinematic dipole amplitude estimation.  
Next-generation galaxy surveys over huge volumes of the Universe will deliver the accurate dipole measurements which allow us better
test our estimation.

\begin{acknowledgments}
I thank Roy Maartens, Chris Clarkson and Dominik J. Schwarz for valuable comments and discussions. 
I acknowledge financial support from the Claude Leon Foundation as well as South African SKA Project and the NRF (South Africa). 
I am grateful for the possibility of performing the 
numerical computations on the Nordrhein-Westfalen state computing cluster at RWTH Aachen.
\end{acknowledgments}

\end{document}